\documentclass[showpacs,PRE]{revtex4-1}
\usepackage{epsfig}
\usepackage{amssymb,amsmath,amsthm,graphics}
\usepackage{color}
\usepackage[colorlinks=true,linkcolor=blue,citecolor=blue]{hyperref}

\newcommand{\braket}[2]{\langle #1|#2\rangle}

\begin{document}

\title{From localized  spot  to the formation of invaginated labyrinth structures in spatially extended systems}
\author{Ignacio Bordeu} 
\affiliation{Departamento de F\'isica, Facultad de Ciencias, Universidad de Chile, Santiago, Chile.}
\author{Marcel G. Clerc}
\email{marcel@dfi.uchile.cl}
\affiliation{Departamento de F\'{i}sica, Facultad de Ciencias F\'isicas y Matem\'aticas,
Universidad de Chile, Casilla 487-3, Santiago, Chile.}
\author{Ren\'{e} Lefever and Mustapha Tlidi}
\email{mtlidi@ulb.ac.be}
\affiliation{D{\'e}partement de Physique, Facult\'e des Sciences, Universit\'{e} Libre de Bruxelles (U.L.B.), 
CP\ 231, Campus Plaine, B-1050 Bruxelles, Belgium.}

\begin{abstract}
The stability of a circular localized spot with respect to azimuthal perturbations is studied in 
in a variational Swift-Hohenberg model equation. The conditions under which the circular shape undergoes an elliptical deformation that transform it into a rod shape structure are analyzed. As it elongates the rod-like structure exhibits a transversal instability that generates an invaginated labyrinth structure which invades all the space available. 
\end{abstract}

\pacs{
%% Editor 05.45.Yv, % nonlinear dynamics of Soliton
05.45.-a, % Nonlinear dynamics,
89.75.Kd  % Pattern formation in complex systems,
}

\maketitle

\section{Introduction}
\noindent
Many spatially extended systems that undergo a symmetry breaking 
instability  close to a second-order critical 
point can be described by a  real order parameter equation in the form of a Swift-Hohenberg type model, which has been derived in various field of nonlinear science such as  hydrodynamics \cite{SH77}, 
chemistry \cite{Hilali}, plant ecology \cite{efeverPlant}, and nonlinear optics \cite{Tlidi93}.  

 A complex Swift-Hohenberg equation was deduced in the context of lasers \cite{Paul, Lega1, Lega2} and optical parametric oscillators  \cite{Longhi}. Moreover,  to describe the nascent optical bistability with transversal effect in nonlinear optical cavities a real approximation has been deduced \cite{TML} from laser equations. This approximation allowed to predict stable localized structures and organized clusters of them \cite{TML}. A detailed derivation of this equation from first principles can be found in Ref. \cite{Paul}. In this work, we show that this real modified Swift-Hohenberg equation (SHE) of the form
\begin{equation} \label{Eq-SH}
 \partial_t u = \eta + \epsilon u - u^3 - \nu \nabla^2 u - \nabla^4 u
\end{equation}
supports a curvature instability on localized structures that leads to an elliptical deformation that produces a rod-like structure. As the time evolution is further increases, the rod-like structure exhibits a transverse undulation and leads to the formation of invaginated structures. This structure is a labyrinthine pattern diefined by an interconnected structure where the field value is high. The outer part or complement to the invaginated structure corresponds to low field value. This behavior occurs far from any pattern forming instability and requires a bistable behavior between homogeneous steady states.
In Eq.~(\ref{Eq-SH}),  $u=u(x,y,t)$ is a real scalar field, $x$ and $y$ are spatial 
coordinates and $t$ is time.
 The parameter $\eta$ represents the external forcing field which brakes the reflection symmetry $u \rightarrow -u$. The bistability parameter is $\epsilon$. The coefficient $\nu$  in front of a diffusive term $\nabla^{2}$ may change the sign and allows the pattern forming to take place  \cite{Tlidi93,CrossHohenberg,hohen,pismenTB,VLM11}.  Depending on the context in which this equation is derived, the physical meaning of the field variable and parameters adopt different meanings, for instance, in cavity nonlinear optics $u(x,y,t)$ corresponds to light field intensity, while parameters $\{ \eta, \epsilon, \nu \}$ are associated with the injection field, the deviations of the cavity field, and cooperativity, respectively \cite{TML}.
 
 For certain range of parameter values, Eq. (\ref{Eq-SH}) exhibits stable circular localized structures, for $\eta<0$ localized structures appear as isolated peaks of the field $u(x,y,t)$, instead, for $\eta>0$ localized structures are holes in the field. These localized structures have a fixed stable radius for each parameter value.
 Curvature instability of localized spot has been experimentally studied or theoretically predicted  in magnetic fluids \cite{Dickstein93}, in  liquid crystals \cite{Oswald1990,Oswald2000}, in reaction-difusion systems \cite{Pearson,LEE,Pismen-2002,Munuzuri,Kaminaga1,Kaminaga2,Kolok,Davis,Muraotv,Schaak98,Pismen-2002,Hayase1,Hayase2}, in plant ecology  \cite{Meron}, in material science \cite{Ren,Nishiura}, in granular  fluid systems and  in  frictional fluids \cite{Sandnes1,Sandnes2}, and nonlinear optics \cite{TVP02}. 
 The fingering instability  of planar fronts leading to the formation of labyrinth structures has been reported by Hagberg et al. \cite{Hagberg}. In this manuscript we shall focus on circular localized states.

\section{Stability of localized spot}

\begin{figure}[b] 
\centering
\includegraphics[width=8.5 cm]{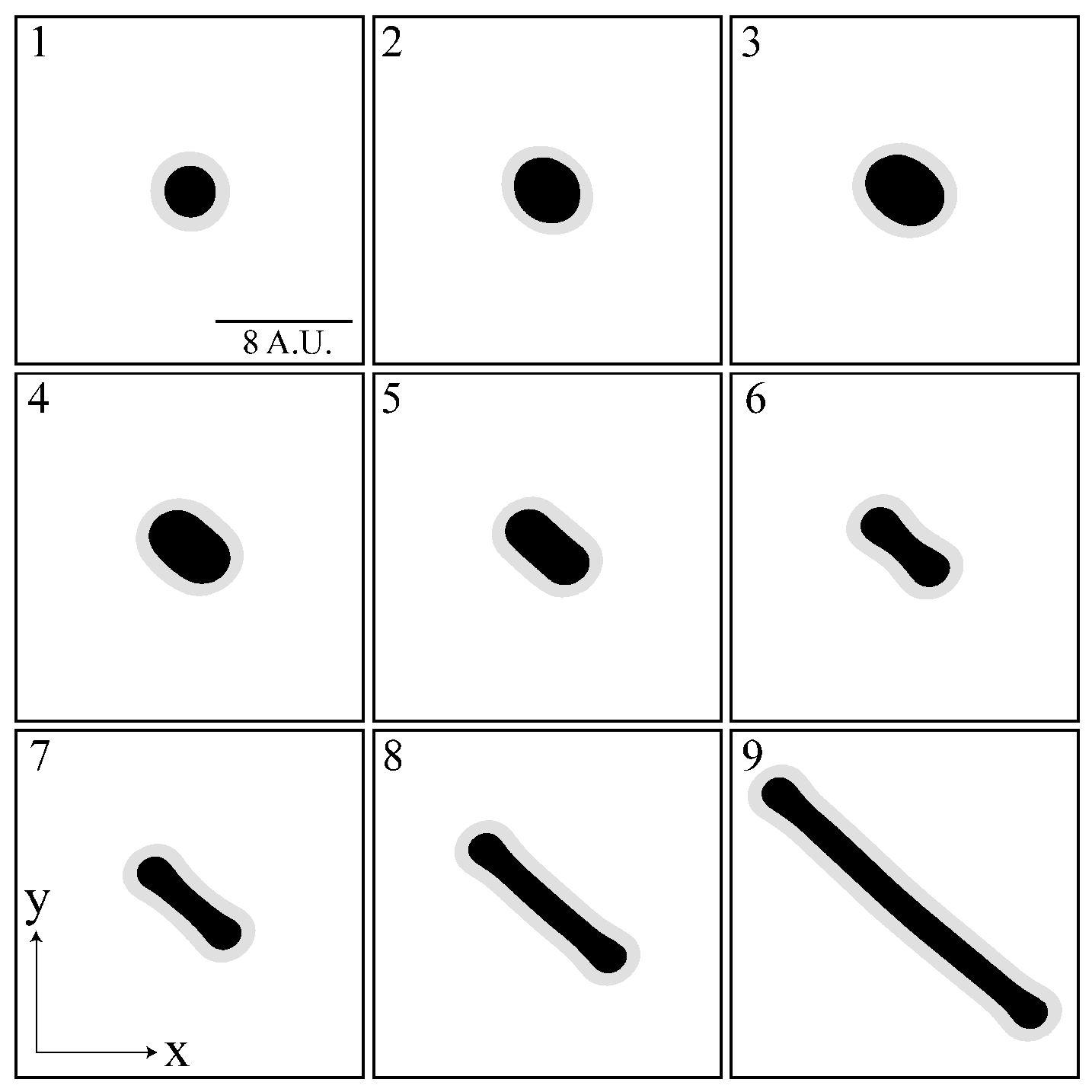}  
\caption{Temporal evolution for; 1) $t=0$; 2) $t=125$; 3) $t=175$; 4) $t=225$; 5) $t=275$; 6) $t=340$; 7) $350$; 8) $360$; 9) $t=400$, of a localized spot into an 
elliptical deformation and then to a rod-like  structure for   Eq.~(\ref{Eq-SH}) with parameters: 
$\eta=-0.065$; $\epsilon=2.45$; $\nu=2.0$. Minima are plain white. The image corresponds to a zoom of $16\times 16$ points of a $512\times 512$ point finite-difference simulation.}
\label{transition}
\end{figure}

\begin{figure}[b] 
\centering
\includegraphics[width=8.5 cm]{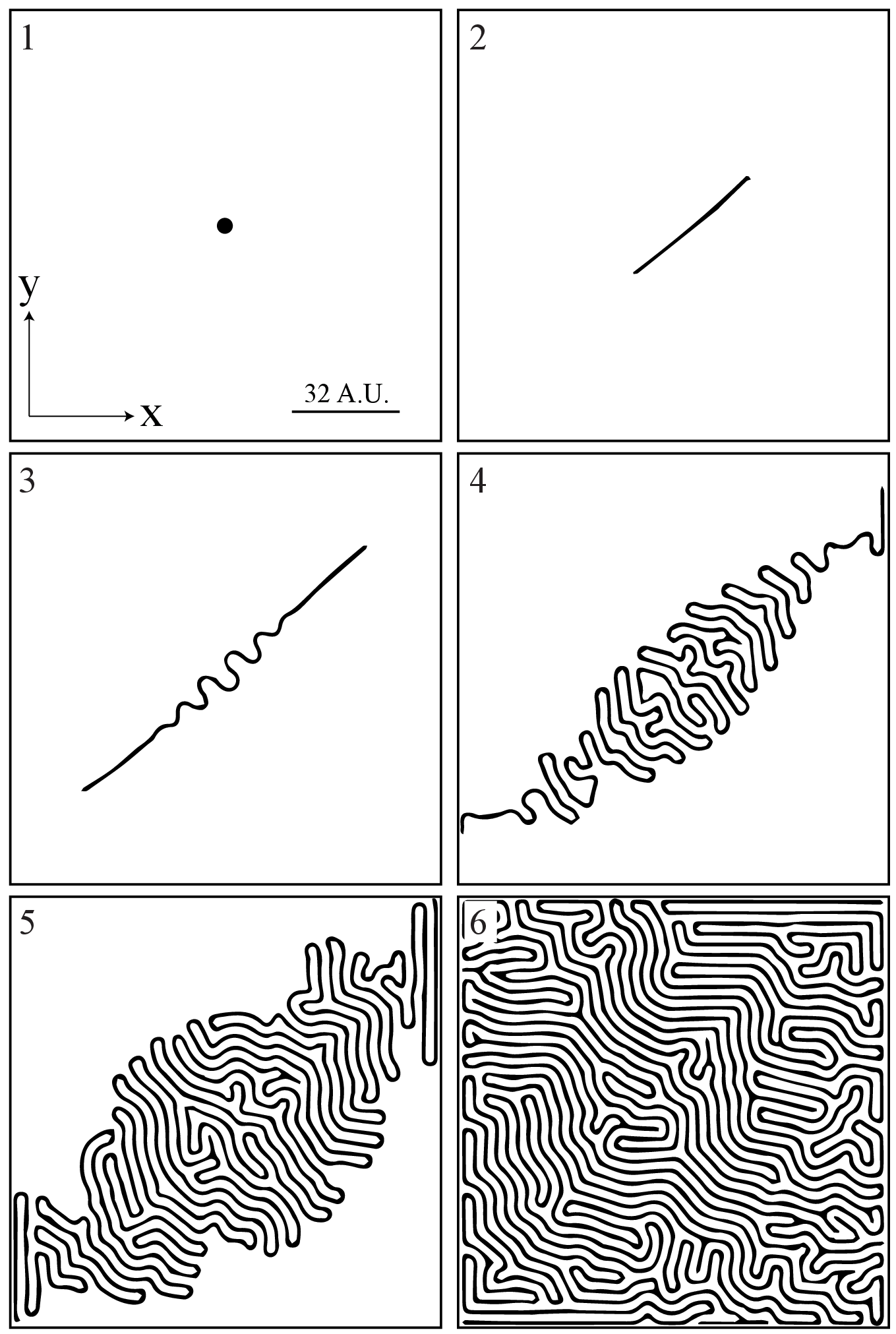}  
\caption{Transition from a single localized spot  to   
invaginated  pattern.
Temporal evolution with
Neumann boundary conditions and with the same parameters as in Fig. \ref{transition}. 1) $t=0$, Localized spot, 2) $t=600$, rod-like structure, 3) $t=1900$, transverse undulation of the rod-like structure 4) $t=2800$, and 5) $t=3700$, localized transient patterns, and 6) $t>15000$, stationary invaginated labyrinth  pattern. Minima are plain white and the mesh integration is $512\times 512$.  Simulation done with finite-difference method}.
\label{numeric}
\end{figure}

Considering fixed parameter values, starting with an azimuthally symmetric localized structure. The structure is perturbed, this perturbation grows radially as shown in Fig. \ref{numeric}.1.  
The circular shape becomes unstable at some critical radius. The elliptical shape elongate into a rod like structures as shown in   Fig.~\ref{transition}.  This  elongation proceeds until  a critical size is reached beyond which 
a transversal instability onset the appearance of fingers near the mid section of the 
structure (see Fig. \ref{numeric}.3). The finger 
continues to  elongate, and the amplitude of  oscillation increases 
(Figs. \ref{numeric}.4 and \ref{numeric}.5).
The dynamic of the system does not saturate and for a long time evolution,  the rod-like structure  
invades the whole space available in ($x$, $y$)-plane as shown in Fig. \ref{numeric}.6. 
This invaginated structure is stationary solutions of the SHE.
The dynamic described previously has been observed in cholesteric liquid crystals under the presence of an external electric field \cite{Oswald1990,Oswald2000}, where an initially circular structure of cholesteric phase suffers from curvature instability, transversal oscillations and develops into an extended labyrinthine structure. The characterization of this dynamic is an open problem.

\begin{figure}[b]
\centering
 \includegraphics[width=8.5 cm]{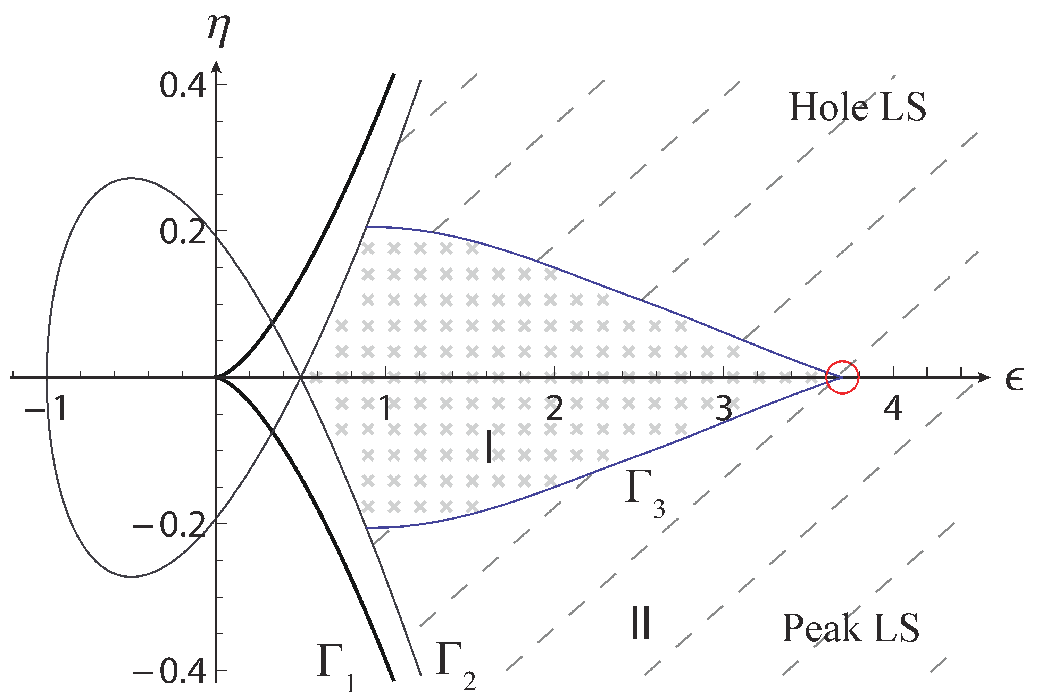}
\caption{(Color online) Bifurcation diagram of  Eq. (\ref{Eq-SH}) in 
($\epsilon$, $\eta$) space for $\nu=2.0$. In $II$-zone (dashed black), stable circular localized structures are observed. In $I$-zone (grey crosses) generation of labyrinthine 
structures are observed from localized structures. }
 \label{diagbif}
\end{figure}

For $\nu=2$, the bifurcation diagram of the model Eq.  (\ref{Eq-SH})  
in the parameter space  
($\epsilon$, $\eta$) is shown in Fig. \ref{diagbif}.  For $\epsilon>0$ the system undergoes a bistable regime between  
homogeneous steady states.  For  $\epsilon<0$,  the system has only one 
homogeneous steady state. The  curve $\Gamma_1$ 
represent the coordinates of the limit points of the bistable curve ($\eta_{\pm} = \pm 2(\epsilon/3 )^{3/2}$).  The threshold associated with a symmetry breaking  
 instability is provided by the curve $\Gamma_2$. The coordinates of the symmetry breaking instabilities thresholds are $\eta_{\pm}=\pm \sqrt{(\nu^2+4\epsilon)/3}(\nu^2-8\epsilon)/24$. Curve $\Gamma_3$, built numerically separates the zone where localized structures are stable, $II$-zone, from the zone where they are unstable, $I$-zone. The  transition from localized structures to  
labyrinthine pattern occurs  when crossing   from $I$-zone to $II$-zone through the  $\Gamma_3$-curves indicated  in 
Fig.  \ref{diagbif}. This transition occurs via fingering instability  at the $\Gamma_3$-curves delimiting the 
parameter domain $I$ and $II$. In the limit of the classical Swift-Hohenberg equation, $\eta=0$, there is no observation of fingering instability, instead the transition from II to I zones, localized structures only grow radially. As a result of boundary conditions destabilization of this structures into labyrinthine structures is observed, this is a size effect phenomenon. Contrary, for $\eta \neq 0$  the transition from II to I zones of a localized spot is affected from curvature instability, giving rise to an unstable rod structure which exhibits transversal oscillations and develops into an extended labyrinthine structure.

In what follows, we first study analytically the stability of a circular localized spot
with respect to azimuthal perturbation. This mode analysis allows us to  
evaluate the threshold above which  the transition from localized spot to a rod-like  structure 
takes place. Then,  we perform the linear stability analyzis of the rod-like 
structure and determine the conditions under which the  transversal oscillations 
occur for the SH equation.


Starting from a solution with 
rotational symmetry (i.e circular localized structure) $u=u_s(r-r_s)$ 
where $r$ is the radial coordinate and $r_s$ characterizes the localized structure radius. 
Then we perturb the solution  $u(X,t) = u_s(X) + W(r,r_s,r_0)$ with $X$ is the relative position
$X\equiv (r-(r_s+r_0 (\theta,t)))$, where $r_0$ 
is the perturbed radius position, which accounts for the interface of the localized structure, 
$\theta$ stands for the angular coordinate, and $W\ll1$ 
are corrections to the circular localized spot. 
Using polar representation of Eq.~(\ref{Eq-SH}), considering the above perturbation and parameters in 
 Eq.~(\ref{Eq-SH}) at linear order in $W$ one obtains
\begin{equation} \label{Eq-SHapp}
\mathcal{L}W=\partial_t r_0 \partial_X u_s +\eta + \epsilon u_s 
- u_s ^3 - \nu \nabla^2 u_s - \nabla^4 u_s,
\end{equation}
 where the lineal operator $
\mathcal{L} \equiv -(\epsilon+3u_s^2-\nu \nabla^2-\nabla^4).
$
 which is a self-adjoint. Assuming that the radius of the localized structure is sufficiently large, 
the operator $\mathcal{L} $ which is explicitly dependent on the radial coordinate 
can be approximated by a homogeneous operator in this radial coordinate

\begin{equation} \label{Eq-SHapp2}
\mathcal{L} \approx -\epsilon - 3u_s^2+ \nu \left( \partial_r^2 + \frac{1}{r_s}\partial_{r} \right)+ \left( \partial_r^4 + \frac{2}{r_s}\partial_r^3 \right),
\end{equation}
this operator possesses a neutral mode, i.e., zero eigenvalue with the eigenfuction  $\partial_X u_s$.
This approach allows us to perform analytical calculations, which are not accessible when the operator is inhomogeneous.
Using this approach the solvability condition (see textbook \cite{pismenTB} and references therein) yields
\begin{equation} 
\label{r0zero}
\partial _t r_0 =- \Delta \frac{1}{r_s^2}\partial_{\theta}^2r_0 
+ \frac{6 \beta}{r^4_s} \partial_{\theta}^2r_0(\partial_\theta r_0)^2 
- \frac{1}{r^4_s} \partial_{\theta}^4r_0 + \frac{2\beta }{r^3_s} (\partial_\theta r_0)^2,
\end{equation}
where
\begin{equation} 
\label{alphabeta}
\beta\equiv\frac{\braket{\partial_{XX} u_s}{\partial_{XX} u_s}}{\braket{\partial_X u_s}{\partial_X u_s}},
\end{equation}
and 
\begin{equation}
\Delta\approx \left( \nu - 2 \beta \right).
\end{equation}
\begin{figure}[b] 
\centering
\includegraphics[width=8.0 cm]{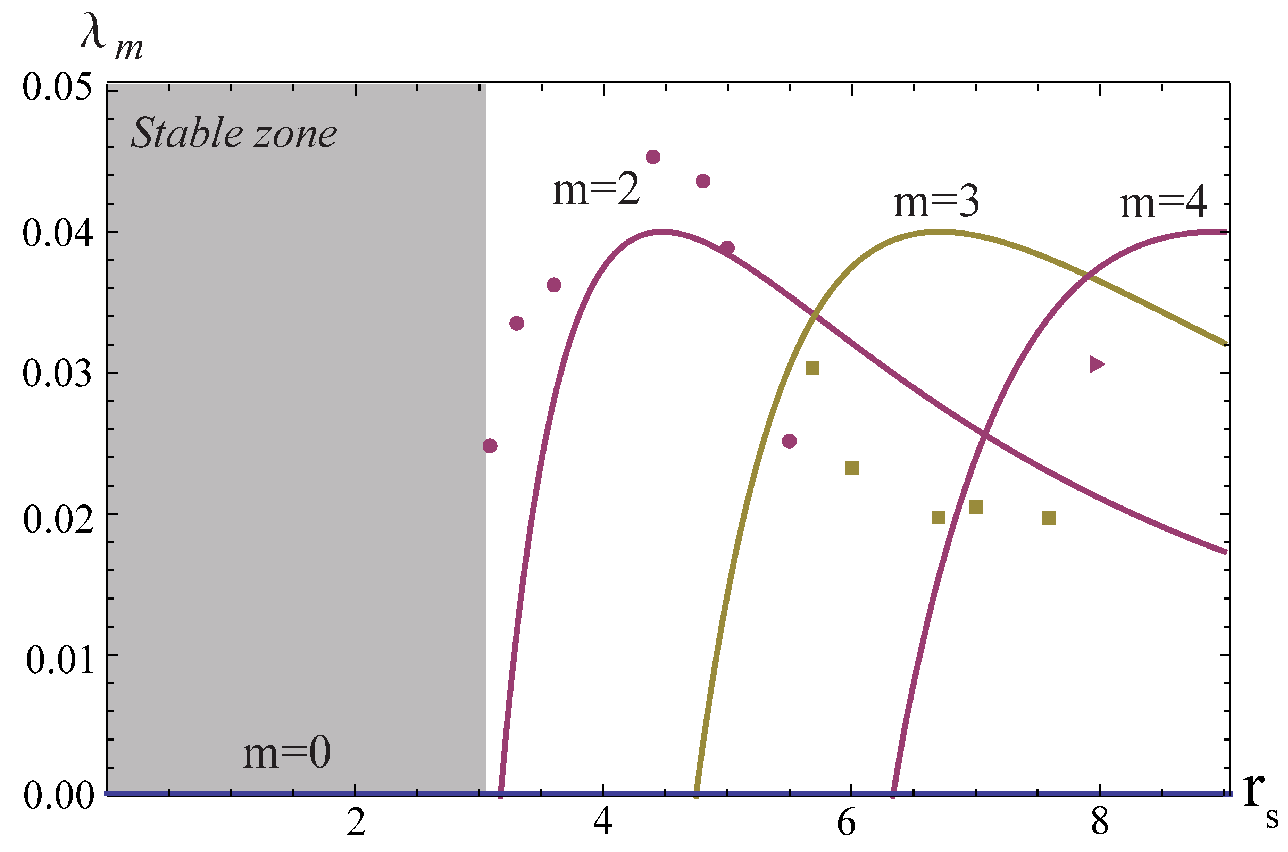}
\caption{(Color online) Lines show approximated theoretical growth rate as function of the localized structure radius
 for $m=0,2,3$ using formula (\ref{lambdaVar}). Points show numerical results for growth rates of modes: $m=2$ (circles), $m=3$ (squares), and $m=4$ (triangle). With parameters: $\eta=-0.065$; $\epsilon=2.45$; $\nu=2$; $dx=0.5$; $dt=0.03$. The insets 
show the different perturbation modes of a localized structure.}
 \label{analisislineal0}
\end{figure}
Then the field $r_0(\theta,t)$  satisfies a nonlinear diffusion equation. The first, the second, the third and the last term   on the right side of Eq.~(\ref{r0zero}) 
account for the linear diffusion, the  non-linear diffusion, the hyper-diffusion and nonlinear advection,
respectively.
When $\Delta < 0$ the localized structure is stable, and for $\Delta>0$, the localized state is unstable as result of the curvature 
instability.
For a better understanding of Eq.~(\ref{r0zero}) one can study the stability of different 
perturbation modes as follows.

 \begin{figure}[t]
\centering
\includegraphics[width=8.0 cm]{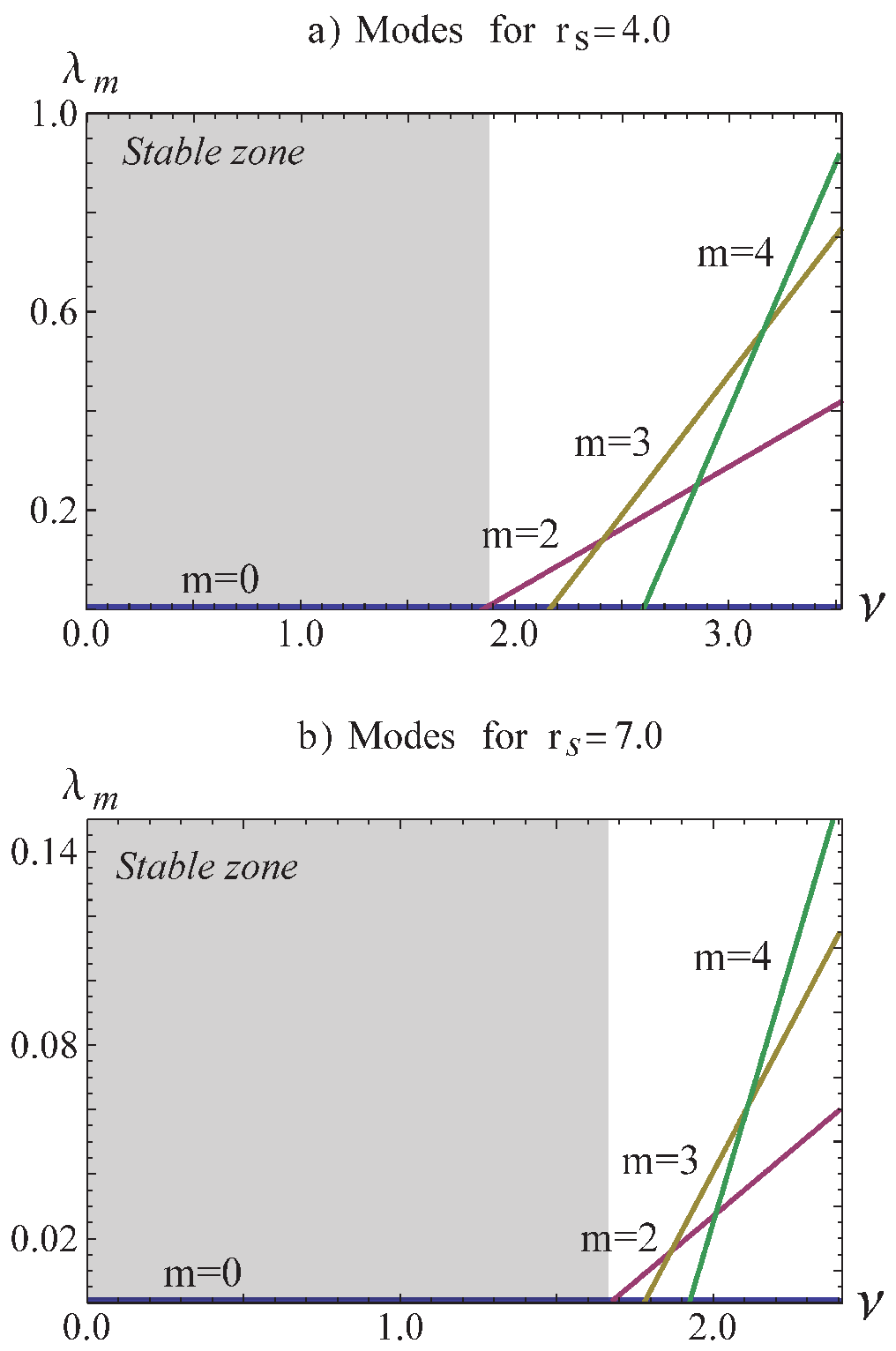}
 \caption{(Color online) Theoretical growth rate as function of $\nu$ using formula~(\ref{lambdaVar}), for parameters: $\eta=-0.065$; $\epsilon=2.45$.
 Emergence of instability modes for fixed radii, for 1) $r_s=4.0$, and 2) $r_s=7.0$.}
\label{rfijo1}
\end{figure}

To study the stability of 2D localized structures, we perturb  $r_0$ with fluctuations of  
the form $r_0(\theta,t)= R_0 Re \text{[} \exp(i m \theta+\lambda_m t) \text{]}$, where, $m$ is integer 
number, $\theta \in \text{[}0,2 \pi \text{[}$, $R_0$ an arbitrary small constant, and $Re$ denotes the real part.
By substituting this perturbations  
in Eq. (\ref{r0zero}) and linearizing in $R_0$, we get the growth rate relation
\begin{equation}
\label{lambdaVar}
\lambda_m (r_s)= ( \nu - 2 \beta ) \left( \frac{m}{r_s} \right)^2 - \left(\frac{m}{r_s} \right)^4.
\end{equation}

 Even though the previous analysis has been made considering a fixed value for the radius of the localized structure, this radius is determined by balance between the interface energy and the energy difference between the homogeneous states. The interface energy and energy difference are proportional to $\nu$ and $\eta$ parameter, respectively.  The radius of the localized structures $r_s$ is proportional to $\nu / \eta$ \cite{coulletLS}, however no analytical expression is available. Hence, the possible radius only lower bounded. This lower bound is determined by the size of the core of the front between homogeneous states. For the sake simplicity, to figure out the role of the growth rate relation we consider that $r_s$ is an independent parameter. Then, for studying the effect of the curvature on the modal instability, one can consider an artificial circular structure of any given radius $r_s$ and evaluate its stability.
Figure~\ref{analisislineal0} shows the growth rate $\lambda_m$ as a function of the radius of circular 2D 
localized structures for different values of angular index $m$. 
For fixed values of all  parameters, the stability of the localized structure is affected  with its radius. 
For small radius, the mode $m=0$ causes radial growth without any change on the structures shape.  
Only modes $m \neq 0$ affect the circular shape of the localized structures. The angular mode  $m=2$ 
becomes unstable ($\lambda_2>0$) and leads to an elliptical deformation of the circular 
shape of the localized structure as shown in Fig \ref{numeric}.2.  This mode has a larger domain of instability 
and an elliptical deformation occurs for $r_0=5.7$ as shown in  Fig. \ref{analisislineal0}. 
Triangular deformation corresponding to the angular index $m=3$ occurs for   $r_0=7.9$.
It has been shown that for fixed parameter values, different modes emerge by means of 
the localized structure's radius increment. Nevertheless, if the radius of the localized structure is fixed, one expects 
that by the variation of a control parameter, lets say $\nu$, one could control the 
emergence of certain unstable modes (cf. Fig.~\ref{rfijo1}). 
In the above results, 
we have  considered that the radius of the localized structure is large, however, this approximation 
is not always valid. Numerical simulations compared to approximate analytical results of the growth rate of localized structures is 
shown in Fig.~\ref{analisislineal0},  numerically, an unstable circular spot of arbitrary radius is generated by hand. From this graphs we conclude 
that the approximate analysis of the $\mathcal{L}$ operator considering $r_s$ as an independent parameter gives a qualitative  description of the observed dynamics. Moreover, the approximated growth rates coincide in magnitude to numerical observations, and the transition radius between unstable modes are consistent.
On the other hand, the stability analysis of  $\mathcal{L}$ is only accessible numerically.

Let us fix the value of the radius to $r_s=4.0$ and to $r_s=7.0$, and 
let now the diffusion coefficient $\nu$ be the control parameter. 
For these values,   the first angular index $m$  mode to become unstable 
is $m=2$ as shown in Fig. \ref{rfijo1}. From this figure we see that when  
increasing $\nu$,  higher order modes become unstable. Comparison 
between the two different values of the radius shows the higher  
unstable modes appear at lower range of values for $\nu$ when increasing the radius $r_0$. 

\section{Transversal instability of rod structures and emergence of labyrinthine patterns}

\begin{figure}[!htbp]
  \centering
 \includegraphics[width=8.0 cm]{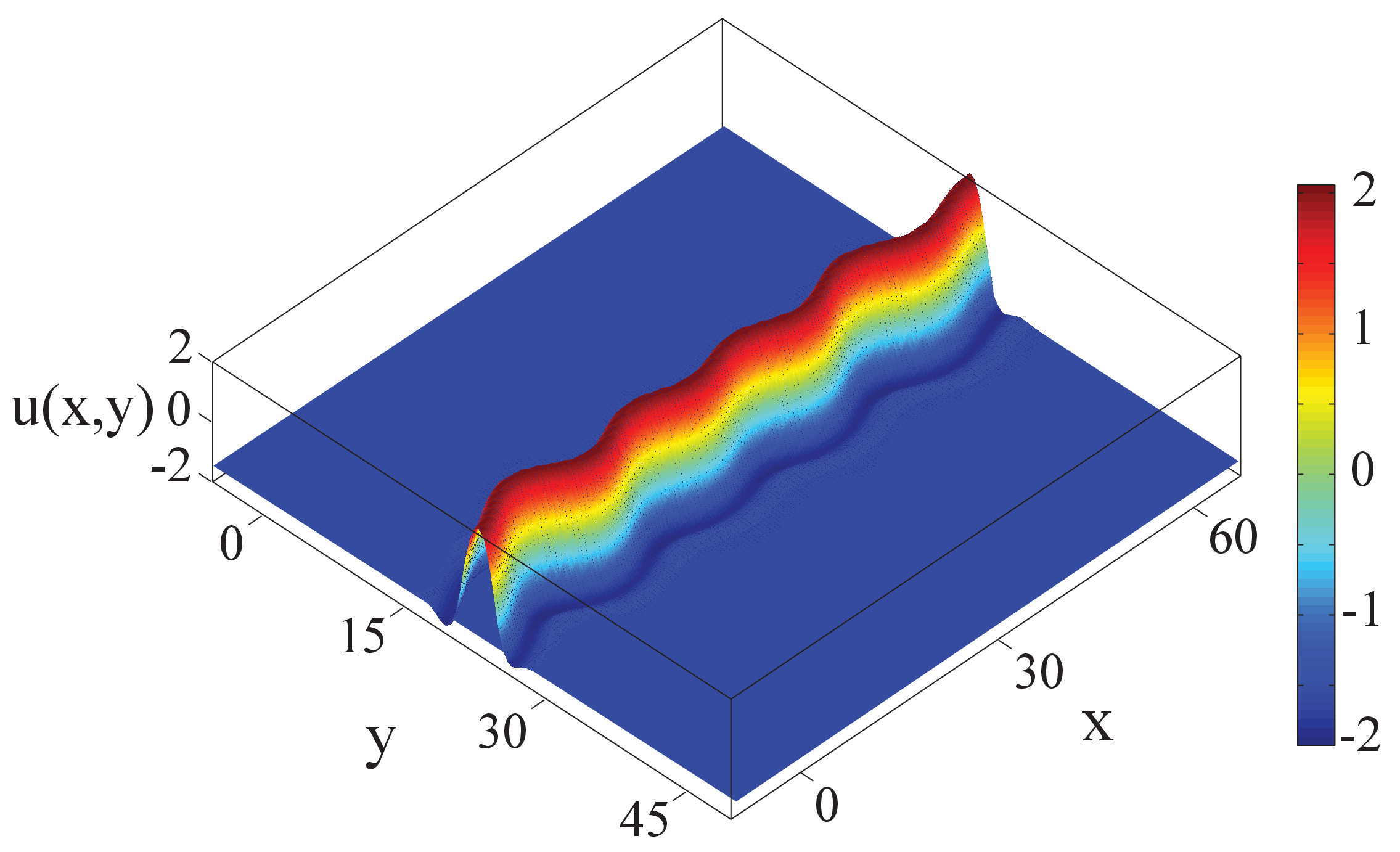}
  \caption{(Color online) Transversal instability of a single infinite stripe of  Eq.~(\ref{Eq-SH}). Image shows a section of a 256$\times$256 point simulation with periodic boundary conditions using a pseudo-spectral code. Parameters: $\eta=-0.065$, $\epsilon=2.45$, $\nu=2$, $dx=0.5$, and $dt=0.03$.}
  \label{inest_dedo}
  \end{figure}

The SHE Eq. (\ref{Eq-SH})  admits a 
single stripe like solution \cite{TLM98,k41}. In order to  evaluate the threshold over which  transversal 
oscillations appear, we perform the  stability analysis of a rod-like structure, by a method similar to the one performed in Ref. \cite{Hagberg}. For this purpose we perturb the single stripe solution as $u = u_f(\boldsymbol{\xi}) + W(\boldsymbol{x},\boldsymbol{X_0})
$
where $u_f$ is the single stripe  solution and $\boldsymbol{\xi} = \boldsymbol{x}-\boldsymbol{X_0} (y,t)$ the relative position, 
$\boldsymbol{X_0}$ is the field that accounts for the shape and evolution 
of the finger, and $W(\boldsymbol{x},\boldsymbol{X_0})<<1$ 
is a non-linear correction of a single stripe. 
Applying this ansatz  in Eq. (\ref{Eq-SH}) at first order in $W$ 
and applying the solvability condition \cite{pismenTB}, the following equation is obtained 
for the dynamic of $\boldsymbol{X_0}$

\begin{equation} 
\label{xzero}
\partial _t X_0 =-\Delta'\partial_{y y}X_0 + 6 \beta' \partial_{y }^2X_0(\partial_y X_0)^2 - \partial_{y }^4 X_0,
\end{equation}
where 
\begin{equation}
\beta' = \frac{\braket{\partial_{\xi \xi} u_f}{\partial_{\xi \xi} u_f}}{\braket{\partial_\xi u_f}{\partial_{\xi} u_f}},  
\text{ and  } \Delta'=( \nu - 2 \beta').
\end{equation}
Thus $\boldsymbol{X_0}$ satisfies a nonlinear diffusion equation. 
This equation describes the dynamics of an interface between two symmetric states \cite{Chevallard,Calisto}.
This model is well known for exhibiting a zigzag instability. Analogously, to the previous section,
when $\Delta < 0$ the single stripe solution is stable, and for $\Delta>0$, the solution is unstable as result of the curvature 
instability. From equation (\ref{xzero}) one expects to observe the single stripe becomes unstable by the appearance 
of an undulation. Figure~\ref{inest_dedo} shows the emergence of this undulated rod-like structure. 
Note that similar dynamical behavior is observed in the propagation of cholesteric finger in 
liquid crystals \cite{Oswald1990,Oswald2000}.
Later, this undulated stripe is replaced by the emergence of facets 
that form a zigzag structure.
However the higher nonlinear terms control the evolution of the single stripe,
then the dynamics of initial zigzag is replaced by the growth of undulations without saturation
as it is depicted in Fig.~\ref{numeric}.4. Therefore, the system displays the emergence of 
a roll-like pattern which is  formed in the middle section of the structure and invades the system 
generating invaginated structure.
(see Fig.~\ref{numeric}.5).

\section{Conclusions}
In this paper we have described the stability of localized spot in a Swift-Hohenberg equation. We first constructed a bifurcation diagram
showing different solutions that appear in different regimes of parameters. Then we 
have shown that the angular index $m=2$ becomes first unstable as consequence of curvature 
instability. This instability leads to an 
elliptical deformation of the  localized spot.   Other type of deformations 
corresponding to $m=3$, $4$ and $5$ occur also in a Swift-Hohenberg equation. 
We have shown also that for a fixed values of the localized spot radius, the angular index 
$m=2$ becomes unstable for small diffusion coefficient while  higher order modes $m>2$ 
become unstable for large values of a diffusion coefficient.   

When angular index 
$m=2$  becomes unstable, the curvature instability of localized spot
produces  an elliptical deformation leading to the formation of rod-like structure.
Subsequently, it generates undulations in the rod-like structure. In the course of time, the space time dynamics leads to the formation of invaginated labyrinth structures. To understand this dynamics, we have performed the 
stability of a single stripe localized structure.
 
 It should be noted that by an offset transformation, $u \rightarrow u + u_0$, where $u_0$ is a 
 constant, Eq. (\ref{Eq-SH}) can be rewritten in such a way that the constant parameter $\eta$ is removed and a quadratic nonlinearity appears. This quadriatic model is equivalent to Eq. (\ref{Eq-SH}). The model with a quadratic nonlinearity has been well studied (see the textbook \cite{pismenTB} and the references therein). This equivalence implies that the results of the present work are also valid for physical systems described by the quadratic model.

\acknowledgements
M.G.C. thanks the financial support of FONDECYT project
1120320. I.B. is supported by CONICYT, Beca de Magister Nacional.  M.T. received support from the Fonds National de la Recherche Scientifique (Belgium).  M.T  acknowledges the financial support of the Interuniversity Attraction Poles program of the Belgian Science Policy Office, under grant IAP 7-35 «photonics@be».

\end{document}